**Title: Ongoing Vaccine and Monoclonal Antibody HIV Prevention Efficacy Trials and Considerations for Sequel Efficacy Trial Designs**

Running Title: Future HIV Vaccine and Monoclonal Antibody Efficacy Trials

Peter B. Gilbert[1,2]

[1]Vaccine and Infectious Disease and Public Health Sciences Divisions, Fred Hutchinson Cancer Research Center, Seattle, Washington, USA

[2]Department of Biostatistics, University of Washington, Seattle, Washington, USA

**Abstract**

Four randomized placebo-controlled efficacy trials of a candidate vaccine or passively infused monoclonal antibody for prevention of HIV-1 infection are underway (HVTN 702 in South African men and women; HVTN 705 in sub-Saharan African women; HVTN 703/HPTN 081 in sub-Saharan African women; HVTN 704/HPTN 085 in U.S., Peruvian, Brazilian, and Swiss men or transgender persons who have sex with men). Several challenges are posed to the optimal design of the sequel efficacy trials, including: (1) how to account for the evolving mosaic of effective prevention interventions that may be part of the trial design or standard of prevention; (2) how to define viable and optimal sequel trial designs depending on the primary efficacy results and secondary "correlates of protection" results of each of the ongoing trials; and (3) how to define the primary objective of sequel efficacy trials if HIV-1 incidence is expected to be very low in all study arms such that a standard trial design has a steep opportunity cost. After summarizing the ongoing trials, I discuss statistical science considerations for sequel efficacy trial designs, both generally and specifically to each trial listed above. One conclusion is that the results of "correlates of protection" analyses, which ascertain how different host immunological markers and HIV-1 viral features impact HIV-1 risk and prevention efficacy, have an important influence on sequel trial design. This influence is especially relevant for the monoclonal antibody trials because of the focused pre-trial hypothesis that potency and coverage of serum neutralization constitutes a surrogate endpoint for HIV-1 infection. Another conclusion is that while assessing prevention efficacy against a counterfactual placebo group is fraught with risks for bias, such analysis is nonetheless important and study designs coupled with analysis methods should be developed to optimize such inferences. I draw a parallel with non-inferiority designs, which are fraught with risks given the necessity of making unverifiable assumptions for interpreting results, but nevertheless have been accepted when a superiority design is not possible and a rigorous/conservative non-inferiority margin is used. In a similar way, counterfactual placebo group efficacy analysis should use rigorous/conservative inference techniques that formally build in a rigorous/conservative margin to potential biases that could occur due to departures from unverifiable assumptions. Because reliability of this approach would require new techniques for verifying that the study cohort experienced substantial exposure to HIV-1, currently it may be appropriate as a secondary objective but not as a primary objective.

**Introduction**

A diverse range of HIV prevention modalities have shown partial efficacy in preventing HIV-1 acquisition in randomized, placebo-controlled phase 2b or phase 3 efficacy trials, or have been shown to be partially effective in observational studies [reviewed in (Bekker, Beyrer, and Quinn 2012) and (Krishnaratne et al. 2016); see also part A in the Appendix]. Set against this complex and evolving background of potential tools for HIV-1 prevention, it remains an open question as to which efficacy trial designs are the most appropriate and optimal for the anti-HIV-1 interventions being tested that are central to the HIV Vaccine Trials Network (HVTN) research agenda – candidate HIV-1 vaccines and passively administered monoclonal broadly neutralizing antibodies (bnAbs). Careful consideration of the suite of available options, along with innovative thinking, will be required to answer this question.

This article has three parts. In part one I summarize the four ongoing HIV prevention trials of two HIV vaccine regimens and one passively infused monoclonal antibody currently being conducted by the HVTN, the latter in partnership with the HIV Prevention Trials Network (HPTN), and in parts two and three I discuss challenges posed to the design and choice of sequel efficacy trials, in a general way and then focusing on specific issues for sequels following the ongoing vaccine and bnAb HIV-1 prevention efficacy trials. For concreteness I restrict attention to two proven intervention modalities – daily oral PrEP and medical male circumcision – and to a single unproven intervention modality currently under efficacy testing – injectable PrEP.

**Part 1: Ongoing HIV prevention efficacy trials in the HVTN**

*Ongoing HIV vaccine efficacy trials.* Figure 1 shows the schemas of the HVTN 702 and HVTN 705/VAC89220HPX2008 (henceforth "HVTN 705") randomized, placebo-controlled efficacy trials and the HIV vaccine regimens. HVTN 702 is funded and sponsored by the U.S. National Institutes of Health NIAID, co-funded by the Bill & Melinda Gates Foundation, and was initiated in Q4 2016. HVTN 705 has the same co-funders, is sponsored by Janssen Vaccines and Prevention B.V., and was initiated in Q4 2017. Part B in the Appendix provides details on the HIV-1 vaccine regimens, referred to as ALVAC/gp120.CC.MF59 and Ad26.Mosaic.gp140.C.alum, respectively.

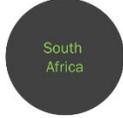

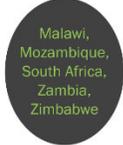

Figure 1. Study schema and HIV-1 vaccine regimen for (A) HVTN 702 and (B) HVTN 705.

The HVTN 702 trial was motivated by several factors including: (1) There is a critical need to develop a subtype C specific HIV vaccine for South Africa given the large epidemic (AVERT 2018); (2) the RV144 efficacy trial of an HIV vaccine regimen in the same regimen class as ALVAC/gp120.CC.MF59 (pox-protein recombinant gp120) previously showed statistically significant partial efficacy (Rerks-Ngarm et al. 2009, Robb et al. 2012); (3) Immune correlates analyses of RV144 demonstrated that certain immune responses to vaccination were correlates of risk of HIV infection and generated hypotheses that these immune responses were also correlates of protection (Haynes et al. 2012), reviews by (Tomaras and Haynes 2014, Corey et al. 2015, Tomaras and Plotkin 2017); and (4) the ALVAC/gp120.CC.MF59 regimen met pre-specified Go-No-Go immunogenicity criteria in the HVTN 100 phase 1 trial based on the correlates of risk that were identified in RV144 (Bekker et al. 2018), including sufficient immune response rates and sufficient variability in immune responses to be able to assess correlates of risk and of protection in HVTN 702.  HVTN 702 has been designed and implemented by the HVTN in collaboration with the Pox Protein Public Private Partnership (P5).

The HVTN 705 trial was also motivated by several factors including: (1) There is a critical need to tackle the problem of global HIV-1 diversity by eliciting greater immune response breadth (Barouch et al. 2010); (2) non-human primate repeated low-dose challenge studies demonstrated efficacy of the Ad26.Mosaic.gp140.C.alum trivalent vaccine regimen to prevent SHIV infection (67% complete protection after 6 intrarectal SHIV challenges) (Barouch et al. 2018); (3) correlates analyses of these challenge trials supported two immune response markers as correlates of protection (Barouch et al. 2018); (4) the Ad26.Mosaic.gp140.C.alum trivalent vaccine regimen met pre-specified Go-No-Go immunogenicity criteria in the APPROACH HIV-V-A004 phase 1/2a clinical trial based on these correlates of risk; and (5) the updated tetravalent vaccine regimen advanced to HVTN 705 showed comparably favorable immunogenicity and cellular response results in the VAC89220HPX2004/HVTN 117/TRAVERSE phase 1/2a trial (Stieh et al. 2018).

The HVTN 702 trial is a hybrid phase 2b/3 trial that is large enough to support a potential licensure application of the vaccine regimen in South Africa, whereas HVTN 705 is a smaller phase 2b test-of-concept trial that provides a less precise characterization of efficacy and is designed to complement a potential Phase 3 trial of a refined version of the regimen. A defining distinguishing mark between phase 3 and 2b trials is testing the null hypothesis of no more than 25% vaccine efficacy vs. testing the null hypothesis of no more than 0% vaccine efficacy, respectively (Rida et al. 1997, Fleming and Richardson 2004, Gilbert 2010). HVTN 702 was designed with licensure potential based on the previous efficacy signal in RV144 such that this trial would provide a second independent efficacy signal that is generally recommended by regulatory agencies to support vaccine licensure (US Food and Drug Administration 1998). In contrast, HVTN 705 is seeking the first evidence for efficacy of an Ad26/gp140 mosaic type of vaccine regimen.

*Ongoing monoclonal antibody HIV prevention efficacy trials.* Figure 2 shows the schema of the two harmonized Antibody Mediated Prevention (AMP) trials (HVTN 704/HPTN 085 and HVTN 703/HPTN 081) and the bnAb regimen description. Both trials are sponsored by the U.S. National Institutes of Health NIAID and were initiated in Q2 2016 (NCT02716675, NCT02568215). The monoclonal antibody being tested in the AMP trials is VRC01, a bnAb that targets the CD4-binding site of the human immunodeficiency virus type 1 (HIV-1) glycoprotein 120 (gp120) (Wu et al. 2010). Participants are randomized to receive ten 8-weekly infusions of VRC01 at 10 mg/kg, VRC01 at 30 mg/kg, or a placebo infusion. HVTN 704/HPTN 085 is being conducted in regions where the circulating HIV-1 viruses are predominantly subtype B and F and enrolled men who have sex with men (MSM) and transgender persons who have sex with men (TG), whereas HVTN 703/HPTN 081 is being conducted in regions where the circulating HIV-1 viruses are predominantly subtype C (although subtypes A and D also circulate in Tanzania, Kenya, and Zimbabwe) and enrolled sexually active women. Further details on the rationale behind the trials and the statistical design of the AMP trials have been described previously (Gilbert et al. 2017).

## Antibody Mediated Prevention Study Schema and Regimen Description

| | | | Infusion Schedule (Weeks) [A = VRC01 infusion; C = Control infusion] | | | | | | | | | | | | |
|---|---|---|---|---|---|---|---|---|---|---|---|---|---|---|---|
| Trial | Treatment | N | 0 | 8 | 16 | 24 | 32 | 40 | 48 | 56 | 64 | 72 | 80* | 88 | 96 | 104† |
| 704/085 Americas/ Lausanne MSM + TG | VRC01 10 mg/kg | 900 | A | A | A | A | A | A | A | A | A | A | | | | |
| | VRC01 30 mg/kg | 900 | A | A | A | A | A | A | A | A | A | A | | | | |
| | Control | 900 | C | C | C | C | C | C | C | C | C | C | | | | |
| | Total | 2700 | | | | | | | | | | | | | | |
| 703/081 Sub-Saharan African women | VRC01 10 mg/kg | 633 | A | A | A | A | A | A | A | A | A | A | | | | |
| | VRC01 30 mg/kg | 633 | A | A | A | A | A | A | A | A | A | A | | | | |
| | Control | 634 | C | C | C | C | C | C | C | C | C | C | | | | |
| | Total | 1900 | | | | | | | | | | | | | | |

Countries: Brazil, Peru, Switzerland, United States (704/085); Botswana, Kenya, Malawi, Mozambique, South Africa, Tanzania, Zimbabwe (703/081).

\* Week 80 is the last study visit for the primary endpoint analysis of prevention efficacy
† Week 104 is the last study visit for assessing safety and tolerability, the tail of VRC01 concentration decline, and a final HIV-1 test

Figure 2. Antibody Mediated Prevention study schema and bnAb regimen description. HIV-1 tests are administered every 4 weeks through Week 80, then at Weeks 88, 96, and 104.

Henceforth "prevention efficacy" (PE) is used to denote either vaccine efficacy or monoclonal antibody efficacy.

*Correlates analyses in the four ongoing prevention efficacy trials.* For each efficacy trial, a key secondary objective assesses various types of "correlates of protection", for understanding how the regimen partially worked (or for characterizing marker distributions if the intervention completely failed to work). It is of interest to assess a variety of types of "correlates of protection," each properly addressed within a specific statistical framework designed for the correlate type – for example survival regression analysis with case-control or case-cohort sampling of immune response markers for studying correlates of risk of HIV-1 acquisition [e.g., (Janes et al. 2017, Breslow et al. 2009)]; machine learning techniques such as Superlearner for studying individual-level signatures of risk [e.g., (Gilbert and Luedtke 2018)]; surrogate endpoint evaluation methods for assessing how well one or more markers adhere to the Prentice (Prentice 1989) definition of a valid surrogate/replacement endpoint for a true clinical endpoint such as HIV-1 infection [e.g., (Kobayashi and Kuroki 2014, Gilbert et al. 2015, Parast, Cai, and Tian 2017)], which includes meta-analysis methods for assessing trial-level validity of surrogate endpoints by assessing the association of PE levels with vaccine/bnAb effects on the putative surrogate (Fleming and DeMets 1996, Fleming and Powers 2012, IOM (Institute of Medicine) 2010)]; principal stratification methods for assessing how prevention efficacy varies over subgroups defined by marker response if assigned to vaccine or bnAb regimen [correlates of prevention efficacy, e.g., (Gilbert and Hudgens 2008,

Gabriel, Sachs, and Gilbert 2015)]; and mediation analysis for seeking insights about mediators/mechanisms of prevention efficacy [(VanderWeele 2015); and see discussion in the Supplement of (Price, Gilbert, and van der Laan 2018)]. Within the term "correlate of protection" I also include virus correlates of prevention efficacy, which are features of exposing HIV-1 viruses, defined based on amino acid sequence characteristics or measured immunological phenotypes, which affect prevention efficacy; sieve analysis methodology assesses whether and how prevention efficacy depends on features of exposing viruses. If some beneficial overall prevention efficacy is demonstrated in an efficacy trial, then it is useful to apply all of these distinct statistical frameworks to the trial data, given that each tackles a distinct scientific question; although first it is prudent to evaluate which frameworks can be expected to provide precise-enough statistical inferences to be worthwhile. In addition, the reviewer reminded me that valid surrogate endpoint evaluation must be based on all relevant efficacy trials, not only the subset of efficacy trials showing some beneficial prevention efficacy; in particular meta-analyses of trial-level validity of surrogate endpoints requires "negative result trials" to be included.

I summarize one particular hypothesized correlate of protection, predicted serum neutralization titer at the time of exposure, which will aid discussion of the issues in the second part. In nonhuman primate (NHP) challenge studies, serum neutralization titer at the time of exposure can be determined by taking serum samples directly prior to challenge and mixing and incubating the challenge virus and Ab/serum prior to performing a neutralization assay (e.g., the TZM-bl target cell assay). However, as this assay is labor-intensive, serum neutralization titer can also be predicted by measuring the serum concentration of the bnAb at the time of challenge (e.g., by ELISA) and dividing this concentration by the $IC_{50}$ of the bnAb against the challenge virus. In a meta-analysis of N=265 NHPs that received a single monoclonal antibody, this day-of-challenge predicted serum neutralizing antibody titer to the challenge SHIV virus was shown to be strongly associated with SHIV transmission (estimated odds ratio 0.012 per 10-fold increase in titer) (Huang et al. 2018), and because 100% of control animals became infected this correlate of risk equates to a correlate of prevention efficacy. Also of note, many licensed vaccines (e.g., yellow fever, inactivated and oral polio vaccines, influenza) have validated neutralization titer surrogate endpoints for a clinically significant infection endpoint [further discussed in (Plotkin 2010, Plotkin et al. 2018)]. While VRC01 was selected for its ability to broadly neutralize most strains of HIV-1, the high genetic diversity of HIV-1 poses a potential challenge to developing such a valid surrogate endpoint, in that individuals are exposed to quasispecies/swarms of HIV-1 virus variants that may frequently contain rare VRC01-resistant variants that potentially could establish infection. Moreover, there are open questions regarding how well the bnAbs reach the site of virus exposure (i.e. how well does serum concentration of the bnAb correlate with the concentration of the bnAb at the relevant mucosal site), whether other immunological functions besides neutralization are needed for protection, what is the relationship between the number of infective HIV-1 virions that need to be neutralized at the exposing site and the number of bnAb molecules present at this site, and whether the particular neutralization assay and readout capture the mechanism of protection (i.e., assay measurement error).

**Part 2: Issues for the Design of Sequel Efficacy Trials (General)**

HVTN 703/HPTN 081 and HVTN 704/HPTN 085 completed enrollment in September and October, respectively, of 2018; both trials are expected to report primary analysis results by Q3 2020. The HVTN 702 and 705 trials are still enrolling and are expected to report primary analysis results by Q2 2022. Part 2 discusses selected issues for the design of sequel efficacy trials in a general way, not considering specific trial designs in terms of specific bnAb or vaccine regimens to evaluate. Part 3 then discusses these issues in terms of specific interventions that may be considered for evaluation in trials following AMP, HVTN 702, and HVTN 705. Part B of the Appendix summarizes HIV-1 vaccine and bnAb regimens currently being studied that could be potentially tested in sequel efficacy trials.

*Appropriateness of a placebo group in a vaccine or monoclonal antibody efficacy trial.* Part D of the Appendix summarizes recommendations from the World Health Organization on the appropriate use of placebo in vaccine trials (Rid et al. 2014, World Health Organization 2013). The central issue is that the control regimen needs to be a version of standard-of-care/prevention in the setting in which the efficacy trial is conducted; this reference point is used in considering study designs below.

*Three types of possible 2-arm sequel efficacy trial designs.* Given the large number of potential prevention interventions that could be constituted by combining different products (e.g., different bnAb cocktails, vaccines, oral PrEP, long-acting injectable PrEP), it behooves researchers to seek promising combinations that may be appealing to certain target populations, have toxicity profiles allowing co-administration, and hold promise for additive or even synergistic prevention effects. If viable, an efficient design would be a 2 by 2 factorial design that allows study of two individual regimens and their combination vs. a shared control regimen, and allows testing for interaction/synergy. We further consider simpler two-arm randomized designs, where we let A denote a new vaccine or bnAb regimen to test after HVTN 702/705 or AMP deliver results, respectively, and let B be another important intervention previously demonstrated to help prevent HIV-1 acquisition that is predicted to be effective in the population being considered for the sequel efficacy trial (e.g., medical male circumcision, oral PrEP, or injectable PrEP if supported by the results of HPTN 083 and HPTN 084). Borrowing the nomenclature of Donnell's article in this issue, Table 1 summarizes three possible general types of efficacy trial designs for studying a new intervention A while accommodating a key proven effective intervention B, all of which would use HIV-1 infection diagnosis as the primary endpoint.

Table 1. Three general types of 2-arm randomized-controlled prevention efficacy trial designs to study intervention A that accommodate a key proven effective prevention intervention B*

| Name of Design | How is B Accounted For? | Randomized Comparison | Population that is Enrolled |
|---|---|---|---|
| Layer | Facilitated access to B for all participants | A vs. A-Placebo | Individuals with a diversity of behaviors related to interest in and receipt of B. To improve efficiency, the design may enrich enrollment of individuals stating an intention to not use B. |
| Compare | B provided to all in one study arm | A + B-Placebo vs. A-Placebo + B | Individuals interested in receiving B |

| Combine | B provided to all participants | A + B vs. A-Placebo + B | Individuals interested in receiving B |

*A = new intervention (Vaccine or bnAb regimen); B could be oral PrEP or injectable PrEP (pending).

A basic question is which design addresses the most important question for the HIV-1 prevention field? The Compare and Combine designs only enroll individuals interested in receiving B, and hence seek to maximize prevention efficacy in this sub-population. In contrast, the Layer design enrolls a broader population including individuals who may state an intention to not receive B, and hence may be advantageous for the long-view ultimate objective to develop a highly efficacious vaccine that can be administered as universally as possible in all sub-populations irrespective of uptake of B, as coverage of vaccination is an important parameter for determining public health impact (Fauci, Folkers, and Marston 2014).

A second basic question is when are the designs scientifically and ethically appropriate? The Layer design is probably only viable if both (1) B is judged to be not nearing the local standard of prevention in the new efficacy trial population; and (2) B is judged to be not too similar to A (in myriad possible factors), as discussed in the next paragraph. The Compare design faces the choice of a superiority vs. non-inferiority trial, which depends on many factors. Generally a superiority design may be preferred if viable, because its results are simple to interpret, whereas in contrast interpretation of non-inferiority design results depends on causal inferences of what is efficacy of B vs. counterfactual placebo, which are sometimes considered under the "constancy assumption" that is based on transportability of previous B vs. placebo designs and is not verifiable empirically based on the trial data (Fleming 2008, Fleming et al. 2011). Yet in some situations only the non-inferiority version of the Compare design seems viable, because a hypothesis that A is superior to B is not credible. For example, non-inferiority designs are well-motivated when A is hypothesized to have efficacy at least close to that of B – but not expected to be superior – yet has other advantages over B such as easier adherence, reduced toxicity, or reduced cost. Two examples are when B is a proven-efficacious vaccine regimen and A is the same vaccine regimen with a simplified schedule, and when B is a proven-efficacious bnAb regimen and A is the same regimen at smaller dose, reducing cost. The Combine design's scientific justification may require a credible hypothesis that A+B is superior to B, which may be better supported if A and B have distinct mechanisms of action. Given a possible Achilles heel of vaccine regimens to not protect in the initial period post first vaccination given the time it takes for protective immunity to accrue, an example of a potential Combine design would provide all participants a regimen designed for "instant protection" (e.g., PrEP or a bnAb regimen) for 3−6 months and randomize participants to a sequence of vaccinations vs. placebo over 6−24 months, and would assess vaccine efficacy in the presence of the "instant protection" intervention (an idea suggested by Scott Hammer, among others). Note that if B were a version of standard-of-care/prevention and it would plausibly improve upon the efficacy of A, then a Combine design of A+B vs. A-placebo + B would be ethically appropriate based on WHO recommendations.

*What factors make two intervention types cluster together in the same class such that a proven-efficacious intervention must be provided (Compare and Combine designs, not Layer)?* If A and B were identical in all material respects, then the Layer design would be absurd, such that its viability depends

on A and B lying in distinct modality classes. Extensive deliberation among diverse stakeholders across different disciplines is important for determining for each specific future efficacy trial design whether A and B are different enough to make the Layer design viable, with some of the intervention factors necessitating comparison between A vs. B for the future study population including: (1) the participant experience in the mode and frequency of delivery; (2) the biological mechanism; (3) the evidence-basis about the strength of prevention efficacy; (4) side effects; (5) local standard of prevention and availability; and (6) personal goals of participants. For (1), medical male circumcision and daily-pill taking (e.g., oral PrEP) are obviously distinct from injections or infusions, a fact undergirding the HVTN 702−705 Layer designs. Differences in participant experiences are more nuanced for active vaccination by injection vs. injectable PrEP vs. passive bnAb administration, where one could argue that injections on the identical schedule of a vaccine or PrEP are in the same class for factor (1), and increasing differences in the frequency of administration widens their separation, at some point reaching a material difference (e.g., 3 vaccinations over 1 year may be considered to lie in a different modality class than 8-weekly injections over 1 year). Related questions for factor (1) are how distinct are injections vs. infusions for the participant experience, and how distinct are bnAb infusions vs. subcutaneous administration? The answers to these questions could vary by local population characteristics such as country or region within sub-Saharan Africa.

For factor (2), different mechanisms of prevention efficacy include antiretroviral-based (oral and injectable PrEP), passive bnAb-based (monoclonal antibodies), active bnAb-based (novel vaccine research), and vaccine-immunity without significant neutralizing activity (Fc effector immunity and T-cell immunity, e.g., the regimens in HVTN 702 and 705). Knowledge of mechanisms of action for A and B has various impacts including on research for developing correlates of prevention efficacy for A and B. For factor (3), if B has been proven to be very highly efficacious in the study population of the future efficacy trial and A has wide uncertainty about its efficacy, then the Compare and Combine designs may address uninteresting questions, and the Layer design may only be interesting if there is a sizable sub-population that is not interested in B (e.g., this is currently the case for a segment of MSM for medical male circumcision and for oral PrEP). For factor (4), if both A and B are exquisitely safe then likely they would be deemed to lie in the same side-effects modality class, but even rare side effects that differ between interventions could place them in different classes given the high bar for safety of prevention products that are given to healthy volunteers (e.g., rare antiretroviral-based side-effects may differ from rare passive bnAb side-effects).

For factor (5), if B is part of the standard of prevention in the future efficacy trial population, or deemed near enough to it, then the Layer design would be inappropriate following normative ethical principles irrespective of the other factors (although these ethical issues may merit additional discussion if a sub-population declines to take B or cannot take B due to some contra-indication). The standard of prevention consists of a set of communications and actions practiced for all potential study participants, which for some components (e.g., medical male circumcision and oral PrEP) would include a recommendation that the participant strongly consider the component as a prevention option given the evidence-basis for efficacy of the component [as of September 2015, WHO recommends that "oral pre-exposure prophylaxis (PrEP) should be offered as an additional prevention choice for people at substantial risk of HIV infection as part of combination HIV prevention approaches" (World Health Organization 2015)]. If B is not nearing the local standard of prevention, then a variety of differences in local standards and availability may be relevant including intervention cost and the clinical management and infrastructure needed for delivery. For factor (6), personal goals of participants may be relevant, for example if a participant desires to contribute to vaccine development and believes in making the research maximally relevant and efficient toward the long-term goal of near-universal vaccination coverage. These considerations highlight the importance of rich engagement of at risk-communities and listening to preferences and concerns about different intervention modalities, and the need for intervention-acceptability research.

*Impact of oral PrEP and medical male circumcision on future vaccine and bnAb efficacy trial design.* Because oral PrEP has been demonstrated to be highly effective to prevent HIV infection in MSM and men and women in serodiscordant partnerships, with multiple studies demonstrating strong evidence for high efficacy in adherent MSM (Hanscom et al. 2016, Grant et al. 2010, Molina et al. 2015, Baeten et al. 2012, McCormack et al. 2016), there was extensive discussion in the design of each of the four ongoing efficacy trials about how to appropriate oral PrEP. For each trial the conclusion was that a Layer design was appropriate, wherein access to oral PrEP is facilitated for all participants. One reason for this choice as compared to the Compare or Combine designs with B = oral PrEP was to answer the scientific question of efficacy in a broad population that would include the diversity of behaviors around oral PrEP use. Another reason was the efficiency advantage of the Layer design, where the ability to conduct the trial at a smaller sample size makes more efficient use of limited resources and aids adherence to distributive justice (remark from Dr. Glenda Gray at the Symposium). For sequel vaccine/bnAb efficacy trials, the field will again wrestle with the question of whether the Layer design is appropriate, which will include deliberation on how the answer depends on the study population of the sequel efficacy trial. In the trial design processes for the two ongoing HVTN efficacy trials including men (HVTN 702 and HVTN 704/HPTN 085), for medical male circumcision there seemed to be easy consensus that a Layer design was appropriate, where all parties accepted that it should be offered as a prevention option to all participants.

*Impact of injectable PrEP on future vaccine and bnAb efficacy trial design.* HPTN 083 and 084 are randomized double-blinded efficacy trials testing long-acting cabotegravir injected every 8 weeks vs. daily oral emtricitabine/tenofovir disoproxil fumarate (FTC/TDV, Truvada), where HPTN 083 is conducted in MSM and transgender women in the Americas (similar to the HVTN 704/HPTN 085 study population) and HPTN 084 is conducted in sub-Saharan African women (similar to the HVTN 705, HVTN 703/HPTN 081, and the female subgroup of the HVTN 702 study populations). HPTN 083/084 are expected to provide primary analysis results in Q3 2021, after the results from the AMP trials are known and before the results from HVTN 702 and 705 are known. In the event that HPTN 083 or HPTN 084 supports efficacy of injectable PrEP in the respective study population, then for a sequel vaccine or bnAb efficacy trial in the same study population, the field will need to decide whether factors (1), (3)–(6) discussed above (among other possible factors) place the vaccine or bnAb regimen in a distinct enough class from injectable PrEP to make the Layer design viable. Such an argument may be easier for a vaccine regimen than a bnAb regimen, given the considerably sparser administration schedule (factor (1)) of vaccination compared to injectable PrEP, and the long-term goal of near-universal vaccination as discussed above. Given ongoing research that is planning for potential sub-cutaneous administration of a bnAb regimen in a future efficacy trial, one question will be how material is the sub-cutaneous vs. infusion vs. injection route. Factor (3) may be particularly influential; for example, if AMP shows low efficacy of VRC01 and HPTN 083 supports very high efficacy of injectable PrEP, then a Compare design of a new bnAb regimen vs. injectable PrEP or a Combine design of a new bnAb regimen vs. placebo both with injectable PrEP may have limited relevance. An exception may be when a new combination bnAb regimen is predicted to have high efficacy similar to that of injectable PrEP, for example based on the information learned from the correlates analyses from AMP combined with preliminary credible evidence of the likely superiority of a new combination bnAb regimen over VRC01 (such as much higher potency, breadth and/or durability). (A "combination bnAb regimen" refers to a pair or triple of individual bnAbs administered as a cocktail, or to a single bispecific or trispecific molecule that was engineered to target two or three epitopes of HIV-1, respectively.)

*Possible sample sizes of a future vaccine or bnAb efficacy trial Layer, Compare, and Combine designs.* Standard sample size calculations are conducted for a 2-study-arm randomized, blinded trial that assesses multiplicative prevention efficacy defined as one minus the hazard ratio comparing HIV-1 incidence between the two study arms under a proportional hazards model. The needed sample size to achieve 90% power to reject the null hypothesis that the hazard ratio is greater than or equal to a fixed null value based on a 0.025-level 1-sided Cox proportional hazards model score test is determined by a simple formula that inputs the total number of HIV infection events pooled across the two study arms and the true prevention efficacy under the alternative hypothesis (Schoenfeld 1983). For simplicity, the formula is applied without accounting for group sequential monitoring of prevention efficacy. For the Layer design, a null PE of 0% and an alternative PE of 50%, or a null PE of 25% and an alternative PE of 70%, are considered. The former may be relevant if the sequel design was a new phase 2b test-of-concept design without a previous efficacy trial establishing some efficacy of the intervention (e.g., a new type of vaccine regimen), and the latter may be relevant for a Phase 3 design (e.g., a sequel to HVTN 705 that showed moderate efficacy). For the Compare and Combine designs, the null hypothesis is equal infection rates and the alternative hypothesis is a 40% lower rate in study arm 1 vs. study arm 2 (i.e., relative prevention efficacy = 0% vs. 40%). Three scenarios of annual HIV-1 incidence in study arm 2 are considered, based on intervention B being oral PrEP (FTC/TDV), with incidences set roughly based on knowledge of oral PrEP efficacy and on scenarios for possible rates of oral PrEP uptake (the three assumed annual incidences of 0.03, 0.015, and 0.003 correspond approximately to assumptions of 0%, 50%, and 90% of person-years at-risk during protective use of oral PrEP FTC/TDV). The results (Table 2) show that the total sample size needs to be 4 to 5 times larger for the Compare and Combine designs compared to the Layer design, which occurs both because of the lower HIV-1 incidence and the smaller prevention efficacy effect size for the Compare and Combine designs. However, because the results on relative power of the Layer design vs. the other designs depends on the assumption about regimen B uptake, the conclusions may change under alternative assumptions. In particular, the calculations shown suppose 67% efficacy of B vs. hypothetical placebo, and, if this efficacy were reduced, then the sample size advantage of the Layer design would be reduced. In addition, the calculations consider the scenario where the clinically relevant relative reduction in risk is independent from the baseline level of risk. Yet, as suggested by the reviewer, as the baseline risk becomes lower, the relative reduction in risk for a new experimental regimen relative to the standard-of-care control regimen likely would need to be greater, in order to justify that it has a favorable benefit-to-risk profile accounting for its complexity of administration, toxicity, and cost. When this factor is operant the sample size advantage of the Layer design would be reduced. However, in this low baseline risk context it may be challenging to devise a regimen A that can plausibly generate the large relative reduction in risk.

Table 2. Sample sizes of some possible sequel vaccine/bnAb regimen prevention efficacy trial designs of the Layer, Compare, and Combine types for studying intervention A accommodating intervention B

| How is B Accommodated? | Randomized Comparison | Null and Altern. PE for HIV-1 through 2 years | Annual HIV-1 Incidence for Study Arm 1:2 (Under Altern. Hypoth.) | Total Sample Size N |
|---|---|---|---|---|
| Layer Design (B available to all participants through facilitated access) | A vs. A-Placebo | H0: PE = 0% vs. H1: PE = 50% | 0.015 : 0.03 | 2,071 |
| | | | 0.0075 : 0.015 | 4,141 |
| | | | 0.0025 : 0.005 | 12,422 |
| | | H0: PE = 25% vs. H1: PE = 70% | 0.009 : 0.03 | 1,369 |
| | | | 0.0045 : 0.015 | 2,737 |
| | | | 0.0015 : 0.005 | 8,211 |
| Compare Design (B provided to all in one arm) | A + B-Placebo vs. A-Placebo + B | H0: RPE** = 0% vs. H1: RPE = 40% | 0.006 : 0.01 | 10,632 |
| | | | 0.003 : 0.005 | 21,264 |
| | | | 0.0015 : 0.0025 | 42,527 |
| Combine Design (B provided to all in both arms) | A + B vs. A-Placebo + B | H0: RPE = 0% vs. H1: RPE = 40% | 0.006 : 0.01 | 10,632 |
| | | | 0.003 : 0.005 | 21,264 |
| | | | 0.0015 : 0.0025 | 42,527 |

*Power for a log-rank test under proportional hazards with no sequential monitoring and 10% random dropout

**RPE = Relative Prevention Efficacy (PE) against HIV-1 infection, for a superiority hypothesis test (as opposed to a non-inferiority hypothesis test)

*Possible primary efficacy objectives as HIV-1 incidence in all study arms becomes small.* A topic at the Symposium was what will be the primary objective of future efficacy trials for which it is anticipated that HIV-1 incidence will be (very) low in all study arms? The gold standard design powers the trial to detect a difference in HIV-1 incidence based on a multiplicative reduction parameter, such that the priority should be on sustaining this design. One approach to accomplishing this goal focuses on unmet need, which may be represented by subpopulations that remain at higher risk of HIV-1 in spite of the advances established by previous efficacy trials. A second approach to sustaining this design would power the efficacy trial to detect a larger relative risk reduction, which may fit needs in settings where risk levels are low due to improving standard-of-care regimens, as discussed above. If these approaches fail such that retaining a multiplicative reduction parameter would require many tens of thousands of participants, alternatives may be considered that consider a new primary objective, which could be assessed with a lower sample size. Two alternative options would change the primary objective: (1) to assess an additive-difference in HIV-1 incidence parameter; (2) to assess efficacy of each study arm vs. a counterfactual placebo group, measuring efficacy either by the multiplicative reduction parameter or by the averted infections ratio parameter (Dunn et al. 2018). Whereas the number of HIV-1 infection events drives power and precision for the multiplicative reduction parameter, such that evaluation of the trial requires sufficiently high incidence, power and precision for the additive-difference parameter can be high even with a very small number of events, as the total amount of person-years at-risk (i.e., "exposure-time") contributes more to power and precision (Uno et al. 2015, Uno et al. 2014). A criticism of using an additive-difference efficacy parameter is difficulty in generalizing results to new settings, because the parameter depends on background/baseline incidence that may differ in a new setting, and the estimate of the parameter could be near zero purely because the study population has little exposure to HIV-1. This could create an incentive to target enrollment of lower risk individuals to increase the chance of finding positive intervention efficacy, the opposite of what is needed when pursuing scientific evidence about how best to address the remaining unmet need. Therefore, justification for use of an additive-difference parameter for the primary analysis may require evidence that the study cohort experiences a substantial amount of exposure to HIV-1; such evidence may be difficult to gather and uncertainty in the quality of evidence could compromise the interpretability and impact of the trial results. One appealing feature of an additive-difference parameter is that under an assumption that incidence is never higher in study arm 1 vs. study arm 2 (in all subgroups both measured and unmeasured), it has a causal attributable risk interpretation as the probability an individual would be HIV-1 infected if assigned study arm 2 but the infection is averted by assignment to study arm 1 (Huang, Gilbert, and Janes 2012). However, until technologies/methods for measuring HIV-1 exposure can be better validated, the disadvantages of the additive-difference parameter may imply it should only be studied in secondary analyses.

Option (2) of relying on a counterfactual placebo group is fraught with potential risks because estimation of prevention efficacy vs. counterfactual placebo group requires causal epidemiological methods that rely on assumptions that are not fully verifiable even if the trial had an infinite sample size. Therefore, it is generally preferable to conduct large randomized-controlled trials and relegate the counterfactual placebo analysis to secondary or exploratory objectives. However, it is interesting that randomized non-inferiority trial designs have been accepted as an appropriate study design in many contexts, including the HIV-1 prevention field (e.g., the HPTN 084 trial), and these designs also rely on the not fully verifiable constancy assumption or related assumption that is typically couched in terms of a counterfactual placebo group (as noted above). The path forward justifying non-inferiority designs in the face of the fraught risk has required both: (a) a determination that a superiority or placebo-controlled design is not viable; and (b) use of a rigorous non-inferiority margin (Fleming 2008, Fleming et al. 2011), which essentially involves setting the definition of success in establishing non-inferiority conservatively, such that if the constancy assumption or related assumption is majorly violated, then the statistical inference about clinically-meaningful non-inferiority still holds.

Motivated by non-inferiority two-arm designs, Dunn and Glidden (2019) in this issue consider two efficacy parameters – standard multiplicative reduction efficacy and the averted infection ratio – both formulated in terms of a parameter $\theta_C$, the proportionate reduction in HIV-1 incidence that would have been observed comparing the control arm regimen with the counterfactual placebo group. They show that both parameters are linear functions of the experimental vs. control regimen rate ratio, such that formulation in terms of $\theta_C$ makes both parameters essentially multiplicative-reduction type parameters (whereas the original formulation of AIR in terms of counterfactual placebo incidence was an additive-difference type parameter with similar issues as noted above); moreover they argue that use of the AIR allows smaller sample sizes. Each of these efficacy parameters is defined in the same way whether used for a non-inferiority design or other type of design, with design type affecting how the user specifies or estimates $\theta_C$, but for any design type it is advisable to estimate it with a conservative lower bound, given the lack of direct data in the new trial for estimating $\theta_C$. Thus the critical requirements (a) and (b) used to justify and conduct non-inferiority primary analysis translate to equivalent critical requirements for justifying and conducting general counterfactual placebo group primary analysis, where (a) may be re-stated as the inability to identify a satisfactory randomized control regimen for the experimental regimen and (b) may be re-stated as use of a conservative lower bound estimate for $\theta_C$. One way to formalize this conservative lower bound would specify rejection of the null hypothesis of no prevention efficacy (vs. counterfactual placebo) in terms of an estimated uncertainty interval that accounts for uncertainty both due to sampling variability and due to partial non-identifiability of the counterfactual prevention efficacy parameter (Vansteelandt et al. 2006, Gilbert and Huang 2016). The reviewer suggested that, whereas in non-inferiority trials reliance on unverifiable assumptions (such as "constancy") is inherently necessary, it does not appear to be inherently necessary to rely on a counterfactual placebo analysis. While I would agree that there may be other approaches to addressing meaningful efficacy objectives when a desirable randomized control regimen cannot be determined, the fact that all of them would require unverifiable assumptions that link to a counterfactual placebo group seems to be the essential issue that makes all of them align with non-inferiority analysis as essentially the same kind of analysis. Yet, it is a priority to identify creative study designs that avoid being stuck with requirement (a), in which case counterfactual placebo group analysis would be a secondary or exploratory analysis, much more desirable than as a primary analysis.

For the gold-standard and alternative primary objective options discussed above, point and confidence interval estimation of HIV-1 incidence over time for each individual study arm is an important secondary objective. Indeed, in efficacy trials of multiple prevention interventions, a result where HIV-1 incidence is confirmed to be very low in each study group (with appropriate uncertainty quantification) could be a highly useful result, as long as there is supportive (and validated) evidence that the study population did experience substantial exposure to HIV-1 (as noted above). A variety of techniques for gaining such evidence were discussed at the Symposium, including measurement of rectal gonorrhea incidence as a marker of HIV-1 exposure (Jeffrey Murray) and use of a "sentinel cohort," which refers to some imperfect but consistently applied technique for measuring HIV-1 incidence (in the larger study population from which the efficacy trial cohort is sampled) over a long period of time spanning a series of efficacy trials and that would allow "calibrating" estimates of incidence in a counterfactual placebo group (Dean Follmann). One idea for a sentinel cohort in sub-Saharan African women is to use the placebo groups in prevention efficacy trials over time that collect a core set of common baseline covariates, which based on HVTN trials could include HVTN 503 and HVTN 503-S that combined went from 2007 to 2014, HVTN 703/HPTN 081 from 2016 to 2020, HVTN 702 from 2016 to 2022, and HVTN 705 from 2017 to 2022.

**Part 3: Issues for the Design of Sequel Efficacy Trials (Specific to AMP, HVTN 702, HVTN 705)**

*Potential follow-up efficacy trial designs to HVTN 704/HPTN 085 and HVTN 703/HPTN 081 (AMP).* I now consider more specifically potential efficacy trial designs that could follow AMP. All of the designs assume a version of standard-of-care/prevention for all participants, i.e., facilitated access to a package of HIV-1 prevention techniques available in the setting of the trial that would include medical male circumcision and oral PrEP, which follows the Layer design concept when the latter two modalities are considered to be the key proven effective intervention B. While the degree of uptake of the different prevention package elements largely affects the required sample size for all of the designs, our goal is not to estimate these specific sample sizes, but instead to articulate a rationale for different specific sequel designs that could appropriately follow AMP, HVTN 702, or HVTN 705 dependent on the outcomes of those trials. I consider four scenarios of possible AMP trial outcomes defined by the primary analysis of overall PE and the secondary analysis of serum neutralization as a correlate of protection/surrogate endpoint: (A) definitive evidence that PE has a clinically relevant beneficial effect size (e.g., lower 95% confidence limit for PE > 25)% and evidence for serum neutralization as a correlate; (B) definitive evidence for beneficial PE as in (A) and no evidence for a correlate; (C) evidence that PE is near zero or low-to-moderate (e.g., with the lower 95% confidence limit near zero) and there is a trend of evidence for a correlate; and (D) evidence that PE is near zero or low-to-moderate as in (C) and there is no evidence for a correlate (Table 3). The reviewer suggested the distinction between definitive evidence of clinically meaningful benefit as in (A)-(B) vs. test-of-concept screening evidence for promising efficacy as in (C)-(D), with implication that a placebo may be inappropriate for a sequel efficacy trial under (A)-(B) and would be appropriate for a sequel efficacy trial under (C)-(D). For simplicity I do not consider possibly discordant outcomes between the two AMP trials that could occur due to acquisition route and/or HIV-1 subtype issues; the arguments given here would apply but may need to be restricted to a particular setting.

Outcome (A) would provide a straightforward rationale to advance new bnAb regimens to an efficacy trial, because the knowledge of the correlate implies that the new regimens are predicted to have superior efficacy compared to VRC01, given their much-improved neutralization potency and coverage of circulating viruses. As noted above a placebo arm may be inappropriate given lack of equipoise, although this conclusion is not definitive, given that VRC01 is not being considered for approval as a clinical product and thus would not be eligible to be part of a version of standard-of-care in the setting of the sequel efficacy trial; if deemed appropriate, adding a third study arm (bnAb placebo) would add scientific value. In addition, a VRC01 control arm, while possibly ethically acceptable, may be scientifically unappealing given that the new bnAb regimens would be predicted to be majorly superior, and the new bnAb regimen may include VRC01 itself or a close variant version of VRC01, thereby minimizing risk that there is something distinct about the VRC01 molecule that is important for protection. Two such close VRC01 variants are being tested in HVTN phase 1 trials – VRC01-LS and VRC07-523LS – which may have half-life, potency and coverage advantages compared to VRC01 (see part B of the Appendix). Given the pipeline of several double and triple bnAb cocktail regimens and a trispecific molecule, an option would be to test two bnAb regimens head-to-head. This trial may require a large sample size if the primary objective assesses relative multiplicative prevention efficacy; e.g., based on Table 2 the total sample size may need to be between 20 and 40 thousand individuals to detect relative prevention efficacy of 40%. However, focusing on subpopulations with unmet need and powering the trial for larger effect sizes would reduce the needed sample size, as noted above.

Under scenario (B), where there is definitive evidence for clinically meaningful prevention efficacy of VRC01 and the data do not reveal any evidence for a serum neutralization correlate, a different rationale may be needed to advance new bnAb regimens to efficacy testing, given that, in part, the newer regimens are more promising than VRC01 based on the putative correlate that was not corroborated in AMP. Option 1 in Table 3 would use a Combine design that compares a bnAb regimen that includes VRC01 (or one of its close variants as noted above) as a component vs. VRC01 alone as the control arm. This design seems to have a clear rationale given that VRC01 was demonstrated to work well but without understanding of how it works. However, a potential downside of Option 1 is that it includes a single bnAb regimen, and, even with the lack of evidence for a serum neutralization correlate, there is another rationale that combination bnAb regimens – even regimens potentially not using VRC01 – may be predicted to have superior prevention efficacy compared to VRC01. This rationale stems from the fact that HIV-1 exposures are with a diverse swarm of virus variants, where many of these exposures may have minor variants in the swarm that are neutralization-resistant to VRC01 but would be sensitive to at least one bnAb in a cocktail bnAb regimen (or bispecific/trispecific). This hypothesis that a cocktail bnAb regimen would be better able to block acquisition of diverse swarm exposure, which is normative for HIV-1 exposure, is suggested by Julg et al.'s NHP experiment that showed that single bnAb regimens failed to block transmission of a two-SHIV-virus challenge where one of the challenge viruses was resistant to the bnAb, but a double bnAb regimen where each challenge virus was neutralized by one of the bnAbs did block transmission (Julg et al. 2017). The analogous hypothesis for a trispecific antibody is supported by the findings of (Xu et al. 2017). Based on this reasoning, Option 2 for the sequel efficacy trial would study two combination bnAb regimens head-to-head, both of which include VRC01 as a component (either VRC01 in a cocktail or a bispecific/trispecific molecule that targets the VRC01 epitope). Option 3 would also study two combination bnAb regimens head-to-head, one of which includes VRC01 as a component and the other not. Option 3 may be the most difficult to justify given that VRC01 was demonstrated to work well, but this option is included because of the extensive ongoing research on combination bnAb regimens that may generate highly promising regimens that do not necessarily include VRC01. If the scenario (A) and (B) trial designs also have a placebo arm, then primary objectives would include assessment of PE of each tested new bnAb regimen vs. placebo; if there is no placebo arm, then a secondary or exploratory objective would assess PE of each tested new bnAb regimen vs. counterfactual placebo. It is also possible that scenario (B) is a false positive result in that serum neutralization is actually a correlate (such that scenario (A) would have been the correctly classified scenario) but the AMP correlates analysis failed to detect it. Such a false positive result would incur cost to miss the straightforward rationale for next steps of research afforded by scenario (A).

Scenario (C) could be a point estimate of prevention efficacy at 10-20% with the lower confidence limit well below zero. In this scenario, even a nonsignificant trend to corroborate the serum neutralization correlates hypothesis could constitute a solid rationale for advancing new bnAb regimens to efficacy testing, where a placebo-controlled design would likely be viable and advantageous. This rationale is analogous to the history of the development of highly effective antiretroviral therapy for treatment of HIV infected patients.  The initial efficacy studies of zidovudine monotherapy showed small treatment effects to potentially improve progression to AIDS and death and a minor treatment effect on potential surrogate endpoints (first CD4+ T-cell counts and later plasma viral load).  Later research established that solo-zidovudine conferred only minor clinical benefit because of rapid development of drug resistance, and later triple cocktail therapy was shown to have a major effect on both progression to clinical AIDS/death and on plasma virus suppression, which was later validated as a surrogate endpoint. A current hypothesis noted above is that, as a single antibody, VRC01 may have low prevention efficacy coping with swarm exposures that frequently contain VRC01-resistant variants. Such variants could potentially break through with similar efficiency as VRC01-sensitive variants, a problem that could be theoretically circumvented by a cocktail or bispecific/trispecific bnAb regimen.  This model for potential failure of a single bnAb that could be repaired by a cocktail or bispecific/trispecific is referred to as the "overwhelming swarm hypothesis," a hypothesis suggested by the Julg et al. and Xu et al. studies (Julg et al. 2017, Xu et al. 2017) as noted above, as well as by theoretical modeling studies (Wagh et al.  2016, Wagh et al. 2018).

For scenario (D), the overwhelming swarm hypothesis may again constitute a rationale to proceed to a design that tests one or more new cocktail or bispecific/trispecific bnAb regimens versus placebo, as for scenario (C).  If the overwhelming swarm hypothesis is not viewed as credible, then additional research into the reasons why the correlates analysis revealed no evidence to corroborate the serum neutralization correlate hypothesis may be important for finding a path to sequel efficacy trials.  One possible reason would be that the AMP correlates analysis had limited statistical precision, such that the analysis was not able to adequately reject the serum neutralization correlate hypothesis.  Another possible reason would be that serum neutralization is a correlate in an underlying mechanistic sense, but the neutralization assay was insensitive to detect it and other assays or assay-readouts would be needed to provide a correlate/surrogate endpoint.  Possibly an expanded exploratory correlates analysis could discover evidence of a new putative surrogate endpoint, which could change scenario (D) to be more like scenario (C).

There is an important nuance to the assessment of correlates in AMP that needs to be considered, especially relevant for a version of scenario (D) where the analysis results on VRC01 prevention efficacy are 'totally flat,' with estimated overall PE near zero for both the lower and higher dose VRC01 arms and for both first and second-halves of infusion intervals.  For such a result, "sieving" could still occur where the breakthrough founder viruses in the VRC01 group are enriched with VRC01-resistant variants as measured by the neutralization assay and amino acid resistance motifs compared to the placebo group. An interpretation of these results would be that VRC01 only allowed resistant variants of the exposing HIV-1 swarm to transmit, but unfortunately these resistant variants could transmit with about the same efficiency as sensitive variants such that the VRC01-selection pressure did not translate to prevention efficacy.  This pattern of observations from AMP would help support the overwhelming swarm hypothesis and therefore constitute a rationale for testing a new bnAb regimen.  Interestingly, this pattern of observations has a precedent in the HVTN 505 preventive HIV-1 vaccine efficacy trial, which showed estimated overall vaccine efficacy near zero and breakthrough founder viruses in the vaccine group enriched with viruses that departed from the vaccine-strain Envelope sequences in CD4 binding site monoclonal antibody contact sites compared to the placebo group (deCamp et al. 2017).

Table 3. Potential AMP-sequel bnAb efficacy trials (with facilitated access to a version of standard-of-care/prevention in the trial setting, including PrEP) under four scenarios of potential AMP-outcomes

| AMP Shows Definitive Evidence for Clinically Meaningful Prevention Efficacy of VRC01* | | | | AMP Supports None-to-Moderate Prevention Efficacy of VRV01* | |
|---|---|---|---|---|---|
| (A) Evidence for a serum neut. correlate of prevention efficacy | (B) No evidence for a correlate | | | (C) Trend of evidence for a correlate | (D) No evidence for a correlate |
| | Option 1 | Option2 | Option 3 | | |
| bnAb regimen** vs. bnAb regimen (Or/Also vs. Placebo?) (Compare) | bnAb reg. w/ VRC01*** vs. VRC01 (vs. Plac?) (Combine) | bnAb reg. w/ VRC01 vs. bnAb reg. w/ VRC01 (vs. Plac?) (Compare) | bnAb reg. w/ VRC01 vs. bnAb reg. w/o VRC01 (vs. Plac?) (Combine) | bnAb regimen vs. Placebo (Layer) | bnAb regimen vs. Placebo (Layer) |
| Straightforward rationale to test new bnAb regimens, because the knowledge of the correlate implies they are predicted to have superior efficacy compared to VRC01. Evidence for a correlate may come from (1) greater PE of higher dose VRC01; (2) decreasing PE against viruses measured to be more neutralization-resistant to VRC01; (3) greater PE against neutralization sensitive viruses during the first 4 weeks post infusions when VRC01 concentrations are highest; (4) VRC01 concentration and neutralization are strong correlates of risk | More nuanced rationale to test new bnAb regimens, because newer regimens are more promising in part due to a putative correlate for which AMP failed to provide corroborating evidence. Does this imply the designs need to add VRC01 itself or a close VRC01 variant as a control arm? Or, include a version of VRC01 in the newly tested bnAb regimen? Or, is there enough uncertainty in the correlates analysis that a new bnAb regimen may be tested that does not include VRC01 as a component? The 'overwhelming swarm hypothesis' may support 'yes' to the last question.<br><br>- Option 1 may have the strongest rationale, because VRC01, known to be efficacious, is included in both study arms.<br><br>- Options 2 and 3 would require a strong rationale for including a bnAb regimen that does not include VRC01. | | | Even a nonsignificant trend supporting the serum neutralization correlates hypothesis may constitute a rationale for testing new bnAb regimens, given that the newer bnAb regimens are majorly superior to VRC01 in their effect on the correlate. The overwhelming swarm hypothesis may also support a sequel trial. | A new rationale for testing new bnAb regimens would be needed – e.g.: (1) The AMP correlates analysis had limited precision, so the serum neutralization correlate hypothesis was not adequately rejected; (2) Neutralization may be a mechanistic correlate, but the neutralization assay or sampled compartment was insensitive to detect it; (3) Serum neutralization is not a correlate, but the new bnAb regimens are qualitatively different from VRC01 (e.g., by their greater coverage of exposing virus swarms). |

*"Definitive efficacy" [(A), (B)] would support using an active bnAb control regimen, not placebo. However, a third arm (bnAb placebo) may still be viable, based on debates of whether VRC01 must be part of the standard-of-care in the setting of the sequel trial, given it would not be approved as a clinical product and would not be available. A placebo arm is appropriate under the complement scenario "None-to-moderate efficacy" [(C), (D)].

**"bnAb regimen" refers to a 2 or 3 bnAb cocktail, or a bi/trispecific molecule, down-selected for efficacy testing based on HVTN/HPTN Phase 1 trials and supplementary research such as non-human primate challenge studies.

***"VRC01" refers to VRC01 itself and to any of its close variants that may be used instead, based on their improved half-life/potency/coverage (e.g., VRC01-LS or VRC07-523LS).

*Potential follow-up efficacy trial designs to HVTN 705.* Our discussion of considerations for a sequel efficacy trial design following HVTN 705 applies for a sequel design following an arbitrary phase 2b intermediate-sized proof-of-concept efficacy trial that is the first efficacy test of an HIV-1 vaccine candidate concept; thus RV144 being followed by HVTN 702 is a prior example. Table 4 summarizes some considerations for a sequel design, for each of four scenarios of HVTN 705 results parallel to those in Table 3. One difference is that AMP has a stronger/more focused hypothesis about a specific correlate of protection/surrogate endpoint than HVTN 705, given that VRC01 was selected as a broadly neutralizing antibody that has known contact sites with HIV-1 Envelope, whereas active vaccination (as with Ad26.Mosaic.gp140.C.alum) has a more comprehensive and complex impact on the immune system. As a result of this difference, HVTN 705 may not be expected to provide as high as degree of evidence for a surrogate endpoint as is potentially achievable with AMP. For scenario (A), there is a straightforward rationale to test a new Ad26/gp140 mosaic type regimen. The correlates from HVTN 705 would imply that a new regimen of this type would be predicted to have superior efficacy compared to Ad26.Mosaic.gp140.C.alum. The study could use the original regimen Ad26.Mosaic.gp140.C.alum as a comparison arm, and potentially add a placebo arm, with viability of a placebo arm depending on similar considerations made for Table 3, and moreover a placebo would be more appealing if PE in HVTN 705 were moderate and/or if the future trial were in a population with different route of HIV-1 exposure than in HVTN 705 (e.g., MSM + TG instead of sub-Saharan African women). If a placebo arm were deemed appropriate a two-arm new regimen vs. placebo may also be an appealing design. For scenario (B), the lack of evidence for a correlate may strengthen a rationale to use Ad26.Mosaic.gp140.C.alum itself as a control arm. As for the bnAb example, for the designs without a placebo arm an important secondary objective may assess PE of each new vaccine regimen vs. counterfactual placebo.

For scenarios (C) and (D), in which the Ad26.Mosaic.gp140.C.alum vaccine regimen fails to prevent acquisition, this type of vaccination approach may be screened out without additional efficacy trials. Alternatively, a hypothesis that a modification of the vaccine regimen (e.g., making the gp140 boost bivalent as discussed above) would improve efficacy over the vaccine regimen tested in HVTN 705, could open paths to efficacy trials of a refined vaccine regimen, which may viably be placebo-controlled. One rationale for this path is somewhat parallel to the overwhelming swarm hypothesis for AMP – if viral sequence diversity is pinpointed as a likely cause of vaccine failure, then a potential remedy may be improving the content of the gp140 to increase cross-reactive breadth, which is a goal of the refined bivalent gp140 boost that is being developed. A trend of evidence for correlates as in (B) could possibly strengthen a case for additional study of an Ad26/gp140 mosaic type vaccine regimen. All in all, the set of considerations for a sequel bnAb efficacy trial are quite similar to those for a sequel vaccine efficacy trial.

Table 4. Some potential HVTN 705-sequel efficacy trials (with facilitated access to a version of standard-of-care/prevention in the trial setting, including PrEP) under four scenarios of potential HVTN 705-outcomes and remarks on their justification (Considerations generic for an initial Phase 2b intermediate-sized efficacy trial of a candidate HIV-1 vaccine)

| HVTN 705 Shows Definitive Evidence for Clinically Meaningful Prevention Efficacy of the Tested Vaccine | | HVTN 705 Supports None-to-Moderate Efficacy | |
|---|---|---|---|
| (A) Trend of evidence for a correlate | (B) No evidence for a correlate | (C) Trend of evidence for a correlate | (D) No evidence for a correlate |
| New Ad26/gp140 regimen* vs. Ad26.Mosaic.gp140.C.alum (Or/Also vs. Placebo?) (Compare) | New Ad26/gp140 regimen* vs. Ad26.Mosaic.gp140.C.alum (Or/Also vs. Placebo?) (Compare) | New Ad26/gp140 regimen* vs. Placebo? (Layer) | New Ad26/gp140 regimen* vs. Placebo? (Layer) |
| There is a straightforward rationale to test a new Ad26/gp140 regimen. The correlates from HVTN 705 imply that a new Ad26/gp140 based regimen is predicted to have superior efficacy compared to Ad26.Mosaic.gp140.C.alum. Similar remarks as in the first footnote of Table 3 apply as to whether a third study arm (placebo) may be viable. Also, a placebo would be more viable if the future trial were in a population with different route of HIV-1 exposure than in HVTN 705. | Does the lack of evidence for a correlate imply the design needs to add Ad26.Mosaic.gp140.C.alum itself as a control arm? Or, does enough uncertainty remain in the HVTN 705 correlates analysis that a new Ad26/gp140 type regimen may be tested without including Ad26/gp140-C? Similar remarks for (B) about whether a placebo arm may be appropriate. | Potential for major improvements of the HIV-1 vaccine regimen such as making the gp140 bivalent with an optimized mosaic gp140 could motivate another efficacy trial, for example under a hypothesis that viral diversity was a cause of vaccine failure. | Similar remarks as for (C), except there may be stronger arguments for continuing development of an Ad26/gp140 mosaic type vaccine regimen. |

*The new Ad26/gp140 regimen would likely include a bivalent gp140 (subtype C + optimized mosaic).

*Potential follow-up efficacy trial designs to HVTN 702.* If the HVTN 702 trial shows efficacy of ALVAC/gp120.CC.MF59, then it may be licensed depending on the magnitude and durability of efficacy. If the correlates analysis shows (very) strong evidence that one of the hypothesized correlates of protection (anti-V2 antibodies, HIV-1 Env-specific polyfunctional CD4+ T cells, antibody-dependent cellular cytotoxicity) is a correlate of protection/surrogate endpoint in HVTN 702, then it may be possible to follow HVTN 702 with immuno-bridging non-clinical endpoint trials of refined vaccine regimens that are simplified in some sense compared to ALVAC/gp120.CC.MF59 (e.g., with a sparser schedule) that use the surrogate as the primary endpoint and access the accelerated approval regulatory mechanism to achieve direct validation of efficacy against HIV-1 infection in post-approval Phase 4 trials (US Food and Drug Administration 2012). Given that a single phase 3 trial is insufficient for validating a surrogate endpoint, direct HIV-1 endpoint trials of refined vaccine regimens would be needed for full regulatory approval, for example through a non-inferiority trial design of the ALVAC/gp120.CC.MF59 regimen vs. the refined regimen.

**Discussion**

The availability of many proven-effective HIV prevention interventions and the myriad options for use of these interventions in the standard of prevention or in prevention efficacy trial design poses challenges to the design of sequel efficacy trials that would follow the four HIV vaccine and monoclonal antibody randomized placebo-controlled efficacy trials that are currently ongoing. I have considered factors that affect which designs may be viable and optimal, including the appropriate use of Layer, Compare, vs. Combine design approaches to accommodation of a key proven effective intervention (such as oral PrEP), the similarity of prevention modalities, the level of efficacy found in the previous trials, the knowledge learned about correlates of protection in the previous trials, differences in study population in the previous and sequel efficacy trials, and different possible primary efficacy objectives in the sequel efficacy trials. Depending on these factors, randomized trials of new vaccine or monoclonal antibody regimens vs. placebo or vs. other vaccine or monoclonal antibody regimens are viable, where Layer designs that provide facilitated access of key proven effective interventions to all study participants are frequently viable, and advantageous by studying the intervention in a broader study population and for efficient use of resources.

The trial designs without placebo groups often require larger sample sizes, in which case an important secondary objective assesses prevention efficacy of each study arm vs. a counterfactual placebo arm. Credibility of such inferences would be improved by availability of validated technology for measuring HIV-1 exposure, or by availability of a sentinel HIV-1 incidence surveillance cohort that produces reliable incidence estimates (Dean Follmann suggested use of such cohorts at the Symposium). Short of such validated technologies, it seems imprudent to use counterfactual placebo prevention efficacy analysis as a primary analysis. That being said, we have noted fundamental similarities of non-inferiority designs with counterfactual placebo prevention efficacy designs, because interpretability of results from both depends on an assumed counterfactual placebo incidence or on other forms of unverifiable assumptions. For non-inferiority designs it is standard practice to use rigorous and conservative non-inferiority margins that allow for uncertainty in estimates of counterfactual placebo incidence. Applying this principle, vaccine and bnAb counterfactual placebo prevention efficacy analyses should use causal inference methods that formally and conservatively account for uncertainty due to unverifiable assumptions. Another conclusion is that the results of "correlates of protection" analyses, which ascertain different aspects of how HIV-1 risk or prevention efficacy depends on host immunological markers and features of HIV-1 viruses measured from infected trial participants, influence optimal choices for study arms. This influence is especially strong for the monoclonal antibody trials because of the well-supported and focused pre-trial hypothesis about a correlate of prevention efficacy.

While research into innovative design plus analysis techniques is needed, it is also important to continue to prioritize gold-standard designs that use a multiplicative-reduction prevention efficacy parameter and that are powered to detect clinically significant and plausibly-detectable differences between randomized study arms, even when the two arms are active arms, each with expected low incidence. Given the large number of new HIV-1 infections that still occur and the huge human and economic costs of these new infections, the high cost of such trials may still be justified given the maximally rigorous and interpretable answers. For all potential sequel efficacy trial designs, in-country regulatory and ethical considerations, community engagement, and partnership with affected communities will be important for appropriate and effective design, for example when deciding the optimal choice of intervention arms, designing education about the evidence of risks and benefits for all available prevention modalities, and working to facilitate access to these options to minimize HIV incidence while balancing the need to meet the in-country's regulatory requirement for registration or licensure.

*Acknowledgments.* I would like to thank the protocol team of the HVTN 702, HVTN 705, HVTN 703/HPTN 081, and HVTN 704/HPTN 085 trials and the many groups and individuals dedicated to planning and conducting these trials, as well as the study participants. I also thank the organizers (Holly Janes, Deborah Donnell, and Martha Nason) and sponsors of the HIV Prevention Efficacy Trial Designs of the Future Symposium (NIH NIAID, Bill and Melinda Gates Foundation, HVTN, HPTN, MTN) as well as Lindsay Carpp for excellent scientific and technical writing assistance. In addition, I thank Glenda Gray, Steven Nijs, Roels Sanne, Carla Truyers, and An Vandebosch for critical review of the manuscript. Lastly, I thank the peer-reviewer Tom Fleming for his insightful comments that improved this work.

**Conflict of interest statement:** The author has no potential conflicts of interest to declare.

**Funding sources:** This work was supported by the National Institute of Allergy and Infectious Disease at the National Institutes of Health [2 R37 AI054165-11 and UM1 AI068635]. The content is solely the responsibility of the author and does not necessarily represent the official views of the National Institutes of Health.

**Meetings where this research was presented:** P.B.G. presented this research at the HIV Prevention Efficacy Trial Designs of the Future Symposium in November 2018 in Seattle, with talk title "Ongoing Vaccine and Monoclonal Antibody Efficacy Trials in the HVTN and Considerations for Sequel Designs."

**Appendix to "Ongoing Vaccine and Monoclonal Antibody HIV Prevention Efficacy Trials and Considerations for Sequel Efficacy Trial Designs" by Peter B. Gilbert**

*A. Brief summary of the HIV-1 prevention modalities that have shown partial efficacy in preventing HIV-1 acquisition in randomized, placebo-controlled phase 2b or phase 3 efficacy trials, or have been shown to be partially effective in observational studies:*

These interventions include male condoms (Weller and Davis 2002, Pinkerton and Abramson 1997), voluntary medical male circumcision (Siegfried et al. 2009, Wamai et al. 2011), and antiretroviral (ARV) drugs for both HIV-infected persons (e.g. treatment as prevention, TasP) (Cohen et al. 2016, Cohen et al. 2011) and HIV-uninfected persons (e.g. pre-exposure prophylaxis, PrEP) (Grant et al. 2010, Molina et al. 2015, Baeten et al. 2012, McCormack et al. 2016). Behavioral interventions such as individual-level risk reduction counseling have not demonstrated efficacy in randomized controlled trials in preventing sexual transmission of HIV [summarized in (Padian et al. 2010)], but have been shown efficacious in reducing incident sexually transmitted infections (Kamb et al. 1998) and in reducing self-reported high-risk behavior [e.g. (Johnson et al. 2008, Lyles et al. 2007)] and are thus also considered part of comprehensive HIV prevention strategies (Coates, Richter, and Caceres 2008). Together, these interventions comprise a broad spectrum of tools whose effectiveness has different degrees of dependence on human behavioral factors such as adherence and risk compensation [see (Padian et al. 2008) for further discussion of this point]. For instance, medical male circumcision is durably efficacious after a single invasive procedure, whereas daily pill-taking (e.g. oral PrEP) is non-invasive and requires daily or near-daily adherence (Donnell et al. 2014, Murnane et al. 2015). HIV prevention remains an area of intense research, and additional types of modalities have demonstrated partial efficacy in phase 3 trials but have not yet been licensed. Two such examples are a dapivirine-containing vaginal ring (Nel et al. 2016, Baeten et al. 2016) and coital dosing of topical PrEP that showed partial efficacy in one trial (Abdool Karim et al. 2010) but none in a follow-up trial (Delany-Moretlwe et al. 2018).

*B. Ongoing HVTN or HVTN/HPTN prevention efficacy trials of HIV-1 vaccine regimens or bnAb regimens (as of April, 2019) and elaborated details on vaccine and bnAb regimens being studied in current or planned clinical studies.*

Antibody-Mediated Prevention (AMP) trials (HVTN 704/HPTN 085 and HVTN 703/HPTN 081): Harmonized, phase 2b randomized, placebo-controlled monoclonal antibody prevention efficacy trials testing the concept that passive infusion of a monoclonal broadly neutralizing antibody (VRC01) can prevent HIV-1 infection over 80 weeks of follow-up post first infusion.

> bnAb (monoclonal broadly neutralizing antibody): An antibody that neutralizes many different genetic variants of HIV-1.

Combination bnAb regimen: A pair or triple of individual bnAbs administered as a cocktail, or a single bispecific or trispecific molecule that was engineered to target two or three epitopes of HIV-1, respectively.

*Elaborated details on bnAb regimens under consideration for HIV-1 prevention efficacy trials.* Two close VRC01 variants are being tested in HVTN phase 1 trials – VRC01-LS and VRC07-523LS – which may have potency advantages compared to VRC01. VRC01-LS is a modified version of VRC01 that harbors two amino acid point mutations (M428L/N434S) shown to increase the antibody's half-life by 4-fold compared to the parent VRC01 antibody and that retains the serum neutralization activity of the parent VRC01 antibody [as demonstrated in a phase 1 study conducted by the Vaccine Research Center (VRC) (Gaudinski et al. 2018)]. VRC07-523LS was derived from a separate clone from the same donor from whom VRC01 was isolated and is thus a close cousin of VRC01; it also harbors the same two point mutations conferring increased plasma half-life, has been found to be 5 to 8-fold more potent than VRC01, and to confer protection to NHPs at a lower plasma concentration than VRC01 and VRC01-LS (Rudicell et al. 2014).

HVTN 702: Phase 2b/3 randomized, placebo-controlled vaccine efficacy trial testing an ALVAC/gp120.CC.MF59 pox-protein HIV-1 vaccine regimen based on the RV144 vaccine regimen for prevention of HIV-1 infection over 2 years of follow-up post first vaccination.

The ALVAC/gp120.CC.MF59 vaccine regimen being tested in HVTN 702 combines a recombinant canarypox vector with HIV subtype C *env*, subtype B *gag*, and *pol* inserts (ALVAC-HIV [vCP2438]) with a recombinant MF59-adjuvanted bivalent HIV-1 subtype C Envelope (Env) glycoprotein (gp) 120 subunit.

HVTN 705/VAC89220HPX2008: Phase 2b randomized, placebo-controlled vaccine efficacy trial testing a bioinformatically-optimized mosaic Ad26.Mosaic.gp140.C.alum HIV-1 vaccine regimen for prevention of HIV-1 infection over 2 years of follow-up post first vaccination.

The Ad26.Mosaic.gp140.C.alum vaccine regimen being tested in HVTN 705 combines Ad26.Mos4.HIV with a recombinant alum-adjuvanted subtype C gp140 subunit, where Ad26.Mos4.HIV is a tetravalent vaccine comprising four pre-mixed recombinant, replication-incompetent serotype 26 adenoviruses, in a 1:1:1:1 virus particle ratio, two encoding bioinformatically optimized Gag-Pol mosaic proteins and two encoding bioinformatically optimized Env mosaic proteins.

*Potential HIV-1 vaccine regimens for sequel efficacy trials.* Myriad types of candidate HIV-1 vaccines are being tested in the HVTN, defined by vector (e.g., DNA, poxvirus, vesicular stomatitis virus), recombinant Env protein (different strains of gp120 and gp140) or virus-like particle, core proteins such as conserved elements Gag, schedule, route, adjuvant, dose, and mRNA. Directly connected to the ALVAC/gp120.CC.MF59 regimen being tested in HVTN 702, the HVTN (with guidance from the P5) has been interrogating whether immunogenicity can be improved by swapping ALVAC or DNA; changing the gp120 insert-strains, dose, or adjuvant; or modifying the combinations, schedule, or order of vaccine product administration. Directly connected to the Ad26.Mosaic.gp140.C.alum regimen being tested in

HVTN 705, the Janssen/HVTN partnership is testing refinements of this regimen, e.g., by making the gp140 boost bivalent by adding a bioinformatically optimized mosaic gp140 strain. The HVTN is also beginning phase 1 trials of novel candidate vaccines designed to make incremental progress toward vaccine regimens that elicit bnAbs, including lineage-based vaccine design that delivers immunogens sequentially to mimic development of bnAbs in natural infection, germline-targeting vaccine design with primes engineered to activate diverse precursors within a bnAb class and booster immunogens being successively more native-like, and immuno-focusing vaccine design that aims to focus responses to one or more particular epitopes (Havenar-Daughton et al. 2018). The research to develop vaccine regimens that elicit bnAbs is closely tied to the passive bnAb immunoprophylaxis research exemplified by AMP, as the AMP correlates of protection objective aims to estimate a bnAb potency threshold of high efficacy that sets a benchmark for the potency of response required by a bnAb vaccine, and the first bnAb phase 1 candidate vaccine trials that are now underway have as primary endpoints the activation of diverse bnAb precursors in the VRC01 class.

*Potential monoclonal bnAb regimens for sequel efficacy trials.* Given the discovery and development of many monoclonal antibodies that neutralize most strains of HIV-1 and that confer protection to NHPs in challenge studies, the HVTN and HPTN are testing several double and triple bnAb cocktail regimens, involving various combinations of bnAbs including VRC07-523LS, PGT121, and PDGM1400 (HVTN 130 / HPTN 089) (Moldt et al. 2012, Rudicell et al. 2014, Sok et al. 2014, Gautam et al. 2016); combinations with an LS-modified PGT121 are in ongoing development now, as well. The HVTN and HPTN also plan to test a Sanofi-Pasteur trispecific bnAb that combines different paratopes in a single molecule (HVTN 129/ HPTN 088). Trispecific broadly neutralizing HIV antibodies have been shown to completely protect rhesus macaques against a mucosal mixed SHIV challenge (Xu et al. 2017). If AMP shows evidence supporting serum neutralization titer as a correlate of protection, then the primary endpoint for ranking and down-selection of one or more bnAb regimens into a sequel efficacy trial could be based on a score that favors regimens with highest predicted serum neutralization titers against exposing viruses in the future efficacy trial. See Part C of the Appendix for one way to define this score.

C. *bnAb regimen score defined based on predicted serum neutralization titer to exposing HIV-1 viruses in a future planned prevention efficacy trial.*

One way to define such a score for an individual bnAb recipient would be the average across a set of potentially exposing viruses in the future efficacy trial of the area under the predicted bnAb regimen serum inhibitory-dilution 80% ($ID_{80}$)-time curve over the future efficacy trial follow-up period. The details of defining such a score are beyond our scope; elements of such a score endpoint for a given bnAb regimen include: (1) Population pharmacokinetics modeling to estimate serum bnAb time-concentration curves for each bnAb regimen recipient in a phase 1 trial, for each individual bnAb in the regimen (Huang et al. 2017); (2) Measurements of neutralization sensitivity of Envelope pseudoviruses to clinical lots of each individual bnAb in the bnAb regimen (inhibitory concentration 80%, $IC_{80}$), where the Envs are selected to represent potentially exposing viruses in the future efficacy trial [e.g., with $IC_{80}$ data accessed at the Los Alamos National Laboratory "Compile, Analyze and Tally NAb Panels" (CATNAP) repository]; and (3) Prediction of serum $ID_{80}$ for the bnAb regimen against each Env virus used in step (2), for each Phase 1 trial participant at each daily time point over a follow-up period. Step (3) would likely use a model for combination bnAb serum $ID_{80}$ as a function of $IC_{80}$ for each individual bnAb and the estimated concentrations of each bnAb at a given time point, for example based on an additivity assumption (Verrier et al. 2001, Kong et al. 2015) and/or the Bliss-Hill model detailed in (Wagh et al. 2016, Wagh et al. 2018).   Such a formula is important for making efficient use of resources because a fully empirical analysis would need serum sampling and serum neutralization testing on a daily grid.

D. *Recommendations from the World Health Organization on the appropriateness of a placebo group in a vaccine or monoclonal antibody efficacy trial.*

We summarize recommendations from the World Health Organization on the appropriate use of placebo in vaccine trials (Rid et al. 2014, World Health Organization 2013).  These authors deem it unacceptable to use a placebo control when a highly efficacious and safe vaccine exists and is currently accessible in the public health system of the country in which the study is planned, as study participants would have unacceptably high risk of experiencing a delay in receiving benefit from the vaccine through the public health system.  The Nuffield Council on Bioethics guidelines state that the use of a placebo control may be acceptable if participants are not deprived of a vaccine they would have otherwise received, but are provided with the standard of prevention/care that is the best available in the country's public health system.  The Council for International Organizations of Medical Sciences, the International Committee on Harmonization, and UNAIDS (Joint United Nations Programme on HIV/AIDS) guidelines state that researchers must take steps to minimize any risks associated with the use of controls. The protocol should explain clearly the scientific justification for using a placebo-controlled design, and specifically address all of the questions whether (1) the study questions (both efficacy and safety) cannot be answered with an active-controlled study design; (2) the risks of delaying an existing efficacious vaccine are adequately minimized or mitigated; (3) the use of a placebo control is justified by the potential public health value of the research; and (4) the research is responsive to local health needs.